\documentclass[prc,twocolumn,showpacs,preprintnumbers,amsmath,amssymb,superscriptaddress,floatfix,nofootinbib]{revtex4}
\usepackage{graphicx}
\usepackage{amsmath}
\usepackage{amsfonts}
\usepackage{amssymb}%

\newcommand{\Slash}[1]{\ooalign{\hfil/\hfil\crcr$#1$}}

\begin{document}

\title{Role of the $N^*(2080)$ in $pp \to pK^+ \Lambda(1520)$ and $\pi^- p \to K^0 \Lambda(1520)$ reactions }
\author{Ju-Jun Xie} \email{xiejujun@impcas.ac.cn}
\affiliation{Institute of Modern Physics, Chinese Academy of Sciences, Lanzhou 730000, China}
\affiliation{State Key Laboratory of Theoretical Physics, Institute of Theoretical Physics, Chinese Academy of Sciences, Beijing 100190,
China}

\author{Bo-Chao Liu} \email{liubc@xjtu.edu.cn} \affiliation{Department of Applied Physics, Xi'an
Jiaotong University, Xi'an, Shanxi 710049, China}
\affiliation{State Key Laboratory of Theoretical Physics, Institute of Theoretical Physics, Chinese Academy of Sciences, Beijing 100190,
China}

\begin{abstract}

We investigate the $\Lambda(1520)$ hadronic production in the $p p
\to p K^+ \Lambda(1520)$ and $\pi^- p \to K^0 \Lambda(1520)$
reactions within the effective Lagrangian method. For $\pi^- p \to
K^0 \Lambda(1520)$ reaction, in addition to the "background"
contributions from $t-$channel $K^*$ exchange, $u-$channel
$\Sigma^+$ exchange, and $s-$channel nucleon pole terms, we also
consider the contribution from the nucleon resonance $N^*(2080)$
(spin-parity $J^P = 3/2^-$), which has significant coupling to
$K\Lambda(1520)$ channel. We show that the inclusion of the
$N^*(2080)$ leads to a fairly good description of the low energy
experimental total cross section data of $\pi^- p \to K^0
\Lambda(1520)$ reaction. From fitting to the experimental data, we
get the $N^*(2080)N\pi$ coupling constant $g_{N^*(2080)N \pi} = 0.14
\pm 0.04$. By using this value and with the assumption that the
excitation of $N^*(2080)$ is due to the $\pi^0$-meson exchanges, we
calculate the total and differential cross sections of $p p \to p
K^+ \Lambda(1520)$ reaction. We also demonstrate that the invariant
mass distribution and the Dalitz Plot provide direct information of
the $\Lambda(1520)$ production, which can be tested by future
experiments.

\end{abstract}

\pacs{13.75.-n.; 14.20.Gk.; 13.30.Eg.} \maketitle

\section{Introduction}

The study of hadron structure and the spectrum of hadron resonance
is one of the most important issues in hadronic physics and is
attracting much attention (see Ref.~\cite{klempt} for a general
review). In past decades, many excited states of baryon were
observed and their properties have been measured~\cite{pdg2012}. For
those nucleon resonances with mass below $2.0$ GeV, most of their
parameters, such as mass, total decay width, decay modes, etc., have
been more or less studied both on experimental and theoretical
sides. However, for the states around or above $2.0$ GeV, our
present knowledge on them is still in its infancy~\cite{pdg2012}.
Moreover there are still many theoretical predictions of "missing
$N^*$ states" around $2.0$ GeV, which have not so far been
observed~\cite{capstick2000}. Since more number of effective degree
of freedoms will induce more predicted number of excited states, the
"missing $N^*$ states" problem is in favor of the diquark
configuration which has less degree of freedom and predicts less
$N^*$ states~\cite{Zou:2008be}. So, studying the nucleon resonances
around or above $2.0$ GeV, not only on experimental side but also on
theoretical side, is interesting and needed.

The associated strangeness production reaction, $p p \to p K^+
\Lambda(1520)$, is interesting. Firstly, this reaction requires the
creation of an $s\bar s$ quark pair. Thus, a thorough and dedicated
study of strangeness production mechanism in this reaction has the
potential to gain a deeper understanding of the interaction among
strange hadrons and also on the nature of baryon resonances.
Secondly, it is a good channel to study the $N^*$ resonances around
$2.0$ GeV which have significant couplings to $K\Lambda(1520)$
channel, because the $K\Lambda(1520)$ is a pure isospin $1/2$
channel and the production threshold of $K\Lambda(1520)$ is about
$2.0$ GeV ($m_K + m_{\Lambda(1520)} \simeq 2.0$ GeV). Thirdly, the
near threshold differential and total cross sections for kaon pair
production in the $p p \to ppK^+K^-$ reaction have been measured by
DISTO Collaboration~\cite{baledisto}, COSY-11
Collaboration~\cite{Quentmeier:2001ec,Winter:2006vd}, and COSY-ANKE
Collaboration~\cite{hartcosy,zychorcosy,maedacosy}. These results
show clear evidence for the excitation and decay of $\phi$ meson
sitting on a smooth $K^+K^-$ background. For the non-$\phi$ kaon
pair production, the role of the low energy $\Lambda$ excited state,
$\Lambda(1405)$, have been studied in Ref.~\cite{geng1405} by using
chiral unitary theory and in Ref.~\cite{xiecolin1405} within an
unified approach using an effective Lagrangian model. Since the
$\Lambda(1405)$ state lies below $K^- p$ threshold, it is expected
to give significant contribution at the energies near the reaction
threshold. However, at higher energies, the next $\Lambda$ excited
state, $\Lambda(1520)$, could be important for the non-$\phi$ kaon
pair production in the $p p \to p K^+ \Lambda(1520) \to pK^+K^-p$
reaction~\cite{Roca:2006pu}. Furthermore, the $p p \to p K^+
\Lambda(1520)$ reaction is also the basic input for the
$\Lambda(1520)$ production in the proton-nucleus
reactions~\cite{Paryev:2010zza}.

In Refs.~\cite{xiejuan,hejun,nam2013}, the contribution from a
nucleon resonance with spin-parity $3/2^-$ and mass around $2.1$ GeV
was studied in the $\Lambda(1520)$ ($\equiv \Lambda^*$) photo- or
electro-production processes. They all found that this nucleon
resonance has a significant coupling to $K\Lambda(1520)$ channel and
plays an important role in these reactions. Before the year of 2012,
this nucleon resonance was filed as a two-star nucleon resonance
$N^*(2080)$ in the Particle Date Group (PDG) review, which is now
named as $N^*(2120)$~\cite{pdg2012}. Even though, in order for
convenience, here after, we still call it as $N^*(2080)$.

In the present work, we study the role of $N^*(2080)$ resonance
($\equiv N^*$) in the $p p \to p K^+ \Lambda(1520)$ and $\pi^- p \to
K^0 \Lambda(1520)$ reactions within the effective Lagrangian method.
Since the information about $N^*(2080)$ resonance is
scarce~\cite{pdg2012} and the knowledge of its properties, like
mass, total decay width, branch ratios, are poor, we take its mass
and total decay width as $2115$ MeV and $254$ MeV, respectively,
which are obtained by fitting them to the experimental data on the
$\gamma p \to K^+\Lambda(1520)$ reaction in Ref.~\cite{xiejuan}. For
the $N^*(2080)K\Lambda(1520)$ coupling constant, we also take the
value that was obtained in our previous work~\cite{xiejuan}.
Finally, from fitting to the experimental data of $\pi^- p \to K^0
\Lambda(1520)$ reaction, we can get the $N^*(2080)N\pi$ coupling
constant, then we study the role of $N^*(2080)$ resonance in the $p
p \to p K^+ \Lambda(1520)$ reaction with the assumption that the
production mechanism is due to the $\pi^0$-meson exchange.

It will be helpful to mention that though the effective Lagrangian
method is a convenient tool to catch the qualitative features of the
reaction processes, it is not consistent with the unitary
requirements, which in principle are important for extracting the
parameters of the nucleon resonances from the analysis of the
experimental data~\cite{Kamano:2009im,Suzuki:2009nj}, especially for
those reactions involving many intermediate couple channels and
three-particle final states~\cite{Kamano:2008gr,Kamano:2011ih}. In
the present work, basing on phenomenological Lagrangians, we only
consider the tree-diagram contributions, in which the unitarity
condition is not ensured and couple channel effects are not taken
into account. However, our model can give a reasonable description
of the experimental data in the considered energy region. Meanwhile,
our calculation offers some important clues for the mechanisms of
the $\pi^-p \to K^0 \Lambda(1520)$ reaction and make a first effort
to study the role of $N^*(2080)$ resonance in relevant reactions.

In the next section, we will give the formalism and ingredients in
our calculation, then numerical results and discussions are given in
Sect. III. A short summary is given in the last section.

\section{Formalism and ingredients}

The combination of effective Lagrangian method and isobar model is
an important theoretical approach in describing the various
processes in resonance production region. In this section, we
introduce the theoretical formalism and ingredients to calculate the
$\Lambda(1520)$ hadronic production in $\pi^- p \to K^0
\Lambda(1520)$ and $p p \to p K^+ \Lambda(1520)$ reactions within
the effective Lagrangian method.

\subsection{Feynman diagrams and interaction Lagrangian densities} \label{feylag}

\begin{figure}[htbp]
\begin{center}
\includegraphics[scale=0.45]{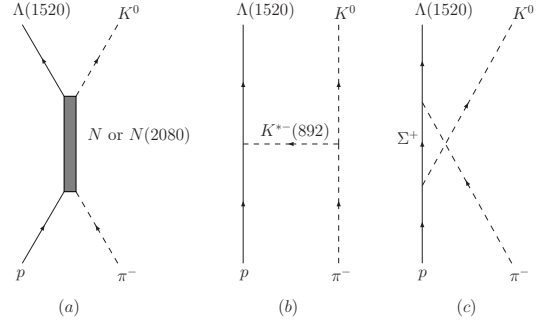}
\caption{Feynman diagrams for $\pi^- p \to K^0 \Lambda(1520)$
reaction. The contributions from $t-$channel $K^*$ exchange,
$u-$channel $\Sigma^+$ exchange, and $s-$channel nucleon pole and
$N^*(2080)$ resonance are considered.} \label{pipdiagram}
\end{center}
\end{figure}

\begin{figure}[htbp]
\begin{center}
\includegraphics[scale=0.45]{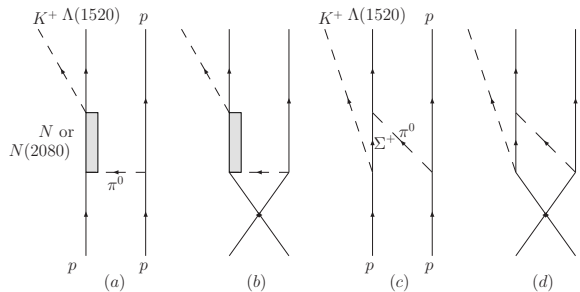}
\caption{Feynman diagrams for $p p \to p K^+ \Lambda(1520)$
reaction. The diagram (a) and (c) show the direct processes, while
(b) and (d) show the exchange processes.} \label{ppdiagram}
\end{center}
\end{figure}

The basic tree level Feynman diagrams for the $\pi^- p \to K^0
\Lambda(1520)$ and $p p \to p K^+ \Lambda(1520)$ reactions are
depicted in Fig.~\ref{pipdiagram} and Fig.~\ref{ppdiagram},
respectively. For the $\pi^- p \to K^0 \Lambda(1520)$ reaction, in
addition to the "background" diagrams, such as $t-$channel $K^*$
exchange (Fig.~\ref{pipdiagram} (b)), $u-$channel $\Sigma^+$
exchange (Fig.~\ref{pipdiagram} (c)), and $s-$channel nucleon pole
diagrams (Fig.~\ref{pipdiagram} (a)), we also consider the
$s-$channel $N^*(2080)$ resonance excitation process
(Fig.~\ref{pipdiagram} (a)). While for the $p p \to p K^+
\Lambda(1520)$ reaction, the $t-$channle $K^*$ exchange process is
neglected since its contribution is small, which will be discussed
below. In Fig.~\ref{ppdiagram}, we show the tree-level Feynman
diagrams for $p p \to p K^+ \Lambda(1520)$ reaction. The diagram
Fig.~\ref{ppdiagram}(a) and Fig.~\ref{ppdiagram}(c) show the direct
processes, while Fig.~\ref{ppdiagram}(b) and Fig.~\ref{ppdiagram}(d)
show the exchange processes.

For the $\pi^- p \to K^0 \Lambda(1520)$ reaction, to compute the
contributions of those terms shown in Fig.~\ref{pipdiagram}, we use
the interaction Lagrangian densities as in
Refs.~\cite{xiejuan,toki,wufq,Mosel,nam2,zouprc03,tsushima,sibi,Saghai,xiezou},
\begin{eqnarray}
{\mathcal L}_{\pi N N}  &=& i g_{\pi N N} \bar{N} \gamma_5 \vec\tau \cdot \vec\pi N, \\ \label{pinn}
\mathcal{L}_{KN\Lambda^*} &=& \frac{g_{KN\Lambda^*}}{m_{K}}\bar{\Lambda}^{*\mu} (\partial_{\mu} K){\gamma}_{5}N\,+{\rm h.c.},  \label{eq:knl} \\
{\mathcal L}_{\pi N N^*}  &=& \frac{g_{\pi N N^*}}{m_{\pi}} \bar{N^*}^{\mu} \gamma_5 (\partial_{\mu}\vec\tau \cdot \vec\pi) N\,+{\rm h.c.},  \label{pinnstar} \\
\mathcal{L}_{K \Lambda^* N^*} &=& \frac{g_1}{m_{K}} \bar{\Lambda}^*_{\mu} \gamma_{5} \gamma_{\alpha} (\partial^{\alpha} K) N^{* \mu} \, +  \nonumber \\
&& \frac{i g_2}{m_{K}^2} \bar{\Lambda}^*_{\mu} \gamma_5 \left (\partial^{\mu} \partial_{\nu} K\right)  N^{*\nu} \,+{\rm h.c.}, \label{eq:eqknstar}
\end{eqnarray}
for the $s-$channel neutron pole and $N^*(2080)$ processes, and
\begin{eqnarray}
{\mathcal L}_{K^* N \Lambda^*}  &=& i g_{K^* N \Lambda^*}
\bar{\Lambda^*}^{\mu} K^*_{\mu} N\,+{\rm h.c.}, \\
\label{kstarnlamstar} \mathcal{L}_{K^*K\pi} &=& g_{K^*K\pi} [\bar{K}
(\partial^{\mu} \vec\tau \cdot
\vec\pi)K^*_{\mu}-(\partial^{\mu}\bar{K})\vec\tau \cdot \vec\pi
K^*_{\mu}] \, \nonumber \\ +{\rm h.c.},  \label{eq:kstarkpi}
\end{eqnarray}
for the $t-$channel $K^*$ exchange process, while
\begin{eqnarray}
{\mathcal L}_{K N \Sigma}  &=& -i g_{K N \Sigma} \bar{N} \gamma_5 K
\vec\tau \cdot \vec\Sigma \,+ {\rm h.c.}, \\ \label{knnsigma}
\mathcal{L}_{\pi \Sigma\Lambda^*} &=&
\frac{g_{\pi\Sigma\Lambda^*}}{m_{\pi}}\bar{\Lambda}^{*\mu}
{\gamma}_{5} (\partial_{\mu} \vec\pi \cdot \vec\Sigma)\,+{\rm h.c.},
\label{eq:pisigmal}
\end{eqnarray}
for the $u-$channel $\Sigma^+$ exchange diagram.

The above Lagrangian densities are also used to study the
contributions of the terms shown in Fig.~\ref{ppdiagram} for $p p
\to p K^+ \Lambda(1520)$ reaction.

It is worth to note that we use the Rarita-Schwinger
formalism~\cite{rarita,nath} to describe the spin $J=3/2$
$\Lambda(1520)$ and $N^*(2080)$ resonances, while the
$\Lambda(1520)$ and $N^*(2080)$ hadronic couplings will be discussed
in the following section.

\subsection{Coupling constants and form factors} \label{ccff}

Firstly, the coupling constant for $\pi NN$ vertex is taken to be
$g_{\pi NN}=13.45$, and the coupling constant $g_{KN\Sigma}$ is
taken as $2.69$ from ${\rm SU}(3)$ symmetry. While the
$N^*(2080)\Lambda(1520)K$ coupling constants $g_{1,2}$ are taken as
the values that we have obtained in Ref.~\cite{xiejuan}, with the
values $g_1=1.4$ and $g_2=5.5$. The $\Lambda^*NK^*$ vertex shown in
Eq.~(\ref{kstarnlamstar}) is predominantly $s-$wave, and the value
of its coupling constant, $g_{K^* N \Lambda^*}$, is $0.5$, which was
obtained and used in Ref.~\cite{toki} (see more details about the
$\Lambda^*NK^*$ couplings in that reference).

Secondly, the coupling constants, $g_{\Lambda^*KN}$, $g_{K^*K\pi}$,
and $g_{\Lambda^*\pi\Sigma}$, are determined from the experimentally
observed partial decay widths of the $K^* \to K\pi$, $\Lambda(1520)
\to KN$, and $\Lambda(1520) \to \pi\Sigma$, respectively. With the
effective interaction Lagrangians described by Eq.~(\ref{eq:knl}),
Eq.~(\ref{eq:kstarkpi}), and Eq.~(\ref{eq:pisigmal}), the partial
decay widths $\Gamma_{\Lambda(1520) \to K N}$, $\Gamma_{K^* \to K
\pi}$, and $\Gamma_{\Lambda(1520) \to \pi \Sigma}$, can be easily
calculated. The coupling constants are related to the partial decay
widths as,

\begin{eqnarray}
\Gamma_{\Lambda(1520) \to K N} &=& \frac{g^2_{\Lambda^*KN}}{6\pi} \frac{|\overrightarrow{p^{{\rm c.m.}}_{N}}|^3(E_N-m_N)}{m^2_K M_{\Lambda^*}}, \label{1520KN} \\
\Gamma_{K^* \to K \pi} &=& \frac{g^2_{K^*K\pi}}{2\pi }\frac{|\overrightarrow{p^{{\rm c.m.}}_{\pi}}|^3}{m^2_{K^*}}, \label{kstarkpi} \\
\Gamma_{\Lambda(1520) \to \pi \Sigma} &=&
\frac{g^2_{\Lambda^*\pi\Sigma}}{4\pi} \frac{|\overrightarrow{p^{{\rm
c.m.}}_{\Sigma}}|^3(E_{\Sigma}-m_{\Sigma})}{m^2_{\pi}
M_{\Lambda^*}}, \label{1520pisigma}
\end{eqnarray}
where
\begin{eqnarray}
E_{N/\Sigma} &=& \frac{M^2_{\Lambda^*}+m^2_{N/\Sigma}-m^2_{K/\pi}}{2M_{\Lambda^*}},
\end{eqnarray}
and
\begin{eqnarray}
|\overrightarrow{p^{{\rm c.m.}}_{N/\Sigma}}| &=& \sqrt{E^2_{N/\Sigma} - m^2_{N/\Sigma}}, \\
|\overrightarrow{p^{{\rm c.m.}}_{\pi}}| &=& \frac{\sqrt{[m^2_{K^*}-(m_K+m_{\pi})^2][m^2_{K^*}-(m_K-m_{\pi})^2]}}{2m_{K^*}}. \nonumber \\
\end{eqnarray}

With mass ($M_{\Lambda^*} = 1519.5$ MeV, $m_{K^*} = 893.1$ MeV),
total decay width ($\Gamma_{\Lambda^*} = 15.6$ MeV, $\Gamma_{K^*} =
49.3$ MeV), and decay branching ratios of $\Lambda(1520)$
[Br($\Lambda^* \to KN$) = $0.45 \pm 0.01$, Br($\Lambda^* \to
\pi\Sigma$) = $0.42 \pm 0.01$] and $K^*$ [Br($K^* \to K\pi$) $\sim
1$], we obtain these coupling constants as listed in
Table~\ref{tab1}.

\begin{table}[htbp]
\begin{center}
\caption{\label{table} Values of the coupling constants required for
the estimation of the $\pi^- p \to K^0 \Lambda(1520)$ and $p p \to p
K^+ \Lambda(1520)$ reactions. These have been estimated from the
decay branching ratios quoted in the PDG book~\cite{pdg2012}, though
it should be noted that these are for all final charged state.}
\begin{tabular}{|ccc|}
\hline Decay modes  & Adopted branching ratios & $g^2/4 \pi$~\footnote{It should be stressed that the partial
decay width determine only the square of the corresponding coupling constants as shown in Eqs.~(\ref{1520KN}, \ref{kstarkpi}, \ref{1520pisigma}),
thus their signs remain uncertain. Predictions from quark model can be used to constrain these signs.
Unfortunately, quark model calculations for these vertices are still sparse. So, in the present calculation, we choose a positive sign for these results.}\\
\hline $\Lambda^* \to KN$  & $0.45$ & $8.77$ \\
       $\Lambda^* \to \pi\Sigma$  & $0.42$ & $0.02$ \\
       $K^* \to K\pi$      & $1.00$  & $0.84$ \\
\hline
\end{tabular} \label{tab1}
\end{center}
\end{table}

Finally, the strong coupling constant $g_{N^*N\pi}$ is a free
parameter, which will be obtained by fitting it to the total cross
sections of $\pi^- p \to K^0\Lambda(1520)$ reaction.

Since the hadrons are not point like particles, we ought to
introduce the compositeness of the hadrons. This is usually achieved
by including the relevant off shell form factors in the amplitudes.
There is no unique theoretical way to introduce the form factors,
and this was discussed at length in the late
nineties~\cite{ohta89,haberzettl98,Davidson:2001rk, Janssen:2001wk}.
We adopt here the common scheme used in many previous works,
\begin{eqnarray}
&& f_i =\frac{\Lambda^4_i}{\Lambda^4_i+(q_i^2-M_i^2)^2},
\quad i= s, t, u, R \\
&&\quad {\rm with} ~~~~~~ \left\{\begin{array}{l}  q_s^2=q_R^2=s, \, q_t^2= t, \, q_u^2= u, \cr
M_s = m_N, \,  M_{\rm R} = M_{N^*}, \cr
 M_{\rm u} = m_{\Sigma}, \cr
M_{\rm t} = m_{K^*},
\end{array}\right. \label{pipff}
\end{eqnarray}
where $s$, $t$ and $u$ are the Lorentz-invariant Mandelstam
variables. In the present calculation, $q_s=q_R=p_1+p_2$,
$q_t=p_1-p_3$, and $q_u=p_4-p_1$ are the 4-momentum of intermediate
nucleon pole and $N^*(2080)$ in the $s-$channel, exchanged $K^*$
meson in the $t-$chanel, and exchanged $\Sigma^+$ in the
$u-$channel, respectively. While $p_1, p_2, p_3$ and $p_4$ are the
4-momenta for $\pi^-$, $p$, $K^0$ and $\Lambda(1520)$, respectively.
Besides, we will consider different cut-off values for the
background and resonant terms, i.e.  $\Lambda_s=\Lambda_t=\Lambda_u
\ne \Lambda_R$.

For $p p \to p K^+ \Lambda(1520)$ reaction, we also need the
relevant off-shell form factors for $\pi NN$ and $\pi N N^*$
vertexes. We take them as
\begin{eqnarray}
F^{NN}_{\pi}(k^2_{\pi}) &=& \frac{\Lambda^2_{\pi}-m_{\pi}^2}{\Lambda^2_{\pi}- k_{\pi}^2}, \label{pinnff} \\
F^{N^* N}_{\pi}(k^2_{\pi}) &=& \frac{\Lambda^{*2}_{\pi}-m_{\pi}^2}{\Lambda^{*2}_{\pi}-k_{\pi}^2}, \label{pinnstarff}
\end{eqnarray}
with $k_{\pi}$ the 4-momentum of the exchanged $\pi$ meson. The
cutoff parameters are taken as $\Lambda_{\pi} = \Lambda^*_{\pi}$ =
1.3 GeV as used in Ref.~\cite{xiezou}.

\subsection{Scattering amplitudes}

For the $\pi^- p \to K^0 \Lambda(1520)$ reaction, with the effective
interaction Lagrangian densities given above, we can easily
construct the invariant scattering amplitudes,

\begin{equation}
-iT_i=\bar u_\mu(p_4,s_{\Lambda^*}) A_i^{\mu} u(p_2,s_p), \label{ti}
\end{equation}
where $u_\mu$ and $u$ are dimensionless Rarita-Schwinger and Dirac
spinors, respectively, while $s_{\Lambda^*}$ and $s_p$ are the spin
polarization variables for final $\Lambda(1520)$ resonance and
initial proton, respectively. To get the scattering amplitude, we
also need the propagators for nucleon and $N^*(2080)$, $K^*$ meson,
and $\Sigma^+$ baryon,
\begin{eqnarray}
G_N(q_s) &=& i \frac{\Slash q_s + m_N}{s-m^2_N}, \\
G^{\mu\nu}_{K^*}(q_t) &=& i \frac{-g^{\mu\nu}+q^{\mu}_tq^{\nu}_t}{t-m^2_{K^*}}, \\
G_{\Sigma}(q_u) &=& i \frac{\Slash q_u + m_{\Sigma}}{u-m^2_{\Sigma}}, \\
G^{\mu\nu}_{N^*}(q_R) &=& i \frac{\Slash q_R + M_{N^*}}{s-M^2_{N^*}+iM_{N^*}\Gamma_{N^*}}P^{\mu\nu},
\end{eqnarray}
and
\begin{eqnarray}
P^{\mu \nu} &=& -g^{\mu \nu} + \frac{1}{3}\gamma^{\mu}\gamma^{\nu} +\frac{2}{3M^2_{N^*}}q_R^{\mu}q_R^{\nu} \nonumber \\
&&
+\frac{1}{3M_{N^*}}(\gamma^{\mu}q_R^{\nu}-\gamma^{\nu}q_R^{\mu}),
\end{eqnarray}
where $M_{N^*}$ and $\Gamma_{N^*}$ are the mass and the total decay width of the $N^*$ resonance.

Then, the reduced  $A_i^{\mu}$ amplitudes in Eq.~(\ref{ti}) can be easily obtained,
\begin{eqnarray}
A_s^{\mu} &=& -i \frac{\sqrt{2}g_{\pi NN}g_{KN \Lambda^*}}{m_K(s-m^2_n)} (\Slash q_s - m_n) \gamma_5 p_3^{\mu}\, f_s, \label{eq:as} \\
A_t^{\mu} &=& -i \frac{\sqrt{2}g_{K^* K\pi}g_{K^* N \Lambda^*}}{t-m^2_{K^*}} (p^{\mu}_1 + p^{\mu}_3 - \nonumber \\
&& \frac{m^2_K-m^2_{\pi}}{m^2_{K^*}}q^{\mu}_t) f_t, \, \label{eq:at} \\
A_u^{\mu} &=& -i \frac{\sqrt{2}g_{\pi \Sigma\Lambda^*}g_{KN \Sigma}}{m_{\pi}(u - m^2_{\Sigma^+})} (\Slash q_u - m_{\Sigma^+}) p_1^{\mu}\, f_u, \label{eq:au} \\
A_{R}^{\mu} &=& i \frac{\sqrt{2}g_{\pi NN^*}}{m_{\pi}m_K D} \Big [ g_1\Slash p_3 (\Slash q_R - M_{N^*})\Big( p^{\mu}_1 - \frac{1}{3}\gamma^{\mu} \Slash p_1 + \nonumber \\
&& \frac{1}{3M_{N^*}}(\gamma^{\mu}q_R \cdot p_1 - q^{\mu}_R\Slash p_1) - \frac{2}{3M^2_{N^*}}q^{\mu}_Rq_R \cdot p_1 \Big) + \nonumber \\
&& \frac{g_2}{m_K} (\Slash q_R - M_{N^*})  p^{\mu}_3 \Big( p_1 \cdot p_3 - \frac{1}{3} \Slash p_3 \Slash p_1 + \nonumber \\
&& \frac{1}{3M_{N^*}}(\Slash p_3 q_R \cdot p_1 - q_R \cdot p_3 \Slash p_1) - \nonumber \\
&& \frac{2}{3M^2_{N^*}}q_R \cdot p_3 q_R \cdot p_1 \Big) \Big] f_R, \label{eq:ar}
\end{eqnarray}
with the sub-indices $s, t, u$ and $R$ stand for the $s-$channel
nucleon pole, $t-$channel $K^*$ exchange, $u-$channel $\Sigma^+$
exchange, and resonance $N^*(2080)$ terms.

For the $pp \to pK^+ \Lambda$ reaction, the full invariant amplitude
in our calculation is composed of four parts corresponding to the
$s-$channel nucleon pole and $N^*(2080)$ resonance, $t-$chanel
$K^*$, and $u-$channel $\Sigma$, which are produced by the
$\pi^0$-meson exchanges, respectively,
\begin{eqnarray}
{\cal M} = \sum_{i = s,~ t,~ u,~R} {\cal M}_{i}. \label{ppamp}
\end{eqnarray}

Each of the above amplitudes can be obtained straightforwardly with
the effective couplings and following the Feynman rules. Here we
give explicitly the amplitude ${\cal M}_s$, as an example,
\begin{eqnarray}
{\cal M}_s & = & \frac{g_{\pi NN}^2 g_{\Lambda^* K N}}{m_K}  F^{N N}_{\pi}(k^2_{\pi}) F^{N^* N}_{\pi}(k^2_{\pi}) F_s(q_N^2) G_{\pi}(k^2_{\pi}) \times
\nonumber\\  && \bar{u}_{\mu} (p_4,s_4)  p^{\mu}_5 \gamma_5 G_N(q_N) u(p_1,s_1)\bar{u}(p_3,s_3) \gamma_5  u(p_2,s_2)  \nonumber\\  &&
  + (\text {exchange term with } p_1 \leftrightarrow p_2), \label{ppampms}
\end{eqnarray}
where $s_i~(i=1,2,3)$ and $p_i~(i=1,2,3)$ represent the spin
projection and 4-momenta of the two initial and one final protons,
respectively. While $p_4$ and $p_5$ are the 4-momenta of the final
$\Lambda(1520)$ and $K^+$ meson, respectively. And $s_4$ stands the
spin projection of $\Lambda(1520)$. In Eq.~(\ref{ppampms}), $k_{\pi}
= p_2-p_3$ and $q_N = p_4+p_5$ stand for the 4-momenta of the
exchanged $\pi$ meson and intermediate nucleon. And
$G_{\pi}(k_{\pi})$ is the pion meson propagator,
\begin{equation}
G_{\pi}(k_{\pi})=\frac{i}{k_{\pi}^2-m^2_{\pi}}.
\end{equation}

\subsection{Cross sections for $\pi^- p \to K^0 \Lambda(1520)$ reaction}

The differential cross section for $\pi^- p \to K^0 \Lambda(1520)$
reaction at center of mass ($\rm c.m.$) frame can be expressed as

\begin{equation}
{d\sigma \over d{\rm cos}\theta}={1\over 32\pi s}{ |\vec{p_3}^{\text{c.m.}}| \over |\vec{p_1}^{\text{c.m.}}|} \left (
{1\over 2}\sum_{s_{\Lambda^*},s_p}|T|^2 \right ), \label{eq:pipdcs}
\end{equation}
where $\theta$ denotes the angle of the outgoing $K^0$ relative to
beam direction in the $\rm c.m.$ frame, while
$\vec{p_1}^{\text{c.m.}}$ and $\vec{p_3}^{\text{c.m.}}$ are the
3-momentum of the initial $\pi^-$ and final $K^0$ mesons. The total
invariant scattering amplitude $T$ is given by,
\begin{equation}
T=T_s + T_t + T_u  + T_R \, .
\end{equation}

From the amplitude, we can easily obtain the total cross sections of
the $\pi^- p \to K^0 \Lambda(1520)$ reaction as functions of the
invariant mass of $\pi^- p$ system. We perform three parameter
($g_{N^*N\pi}$, $\Lambda_s=\Lambda_t=\Lambda_u$ and $\Lambda_R$)
$\chi^2-$fits to the experimental data~\cite{pipdata} on total cross
sections for $\pi^- p \to K^0 \Lambda(1520)$ reaction. There is a
total of $12$ data points below $\sqrt{s}=3.1$ GeV.

\begin{figure}[htbp]
\begin{center}
\includegraphics[scale=0.4]{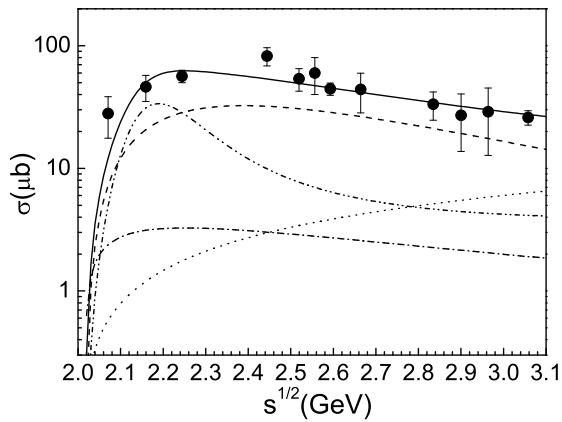}
\caption{Total cross sections vs the invariant mass
$s^{\frac{1}{2}}$ for $\pi^- p \to K^0 \Lambda(1520)$ reaction. The
experimental data are from Ref.~\cite{pipdata}. The curves are the
contributions from $s-$channel nucleon pole term (dashed),
$t-$channel $K^*$ term (dotted), $u-$channel $\Sigma^+$ term
(dash-dotted), $s-$channel $N^*(2080)$ term (dash-dot-dotted) and
the total contributions of them (solid).} \label{piptcs}
\end{center}
\end{figure}

The fitted parameters are: $g_{N^*N\pi} = 0.14 \pm 0.04,
\Lambda_s=\Lambda_t=\Lambda_u=0.89 \pm 0.05$ and $\Lambda_R=0.91 \pm
0.03$. The resultant $\chi^2/dof$ is $1.1$. The best fitting results
for the total cross sections are shown in Fig.~\ref{piptcs},
comparing with the data. The solid lines represent the full results,
while the contributions from the $s-$channel nucleon pole,
$t-$channel $K^*$ exchange, $u-$channel $\Sigma^+$ and $s-$channel
$N^*(2080)$ terms are shown by the dashed, dotted, dot-dashed, and
dash-dot-dotted lines, respectively. From Fig.~\ref{piptcs}, one can
see that we can describe the experimental data of total cross
sections quite well, while the $s-$channel nucleon pole and
$N^*(2080)$ resonance and also the $u-$channel $\Sigma^+$ exchange
give the dominant contributions below $\sqrt{s}=2.4$ GeV. The
$t-$channel $K^*$ exchange diagram gives the minor contribution.

\begin{figure}[htbp]
\begin{center}
\includegraphics[scale=0.4]{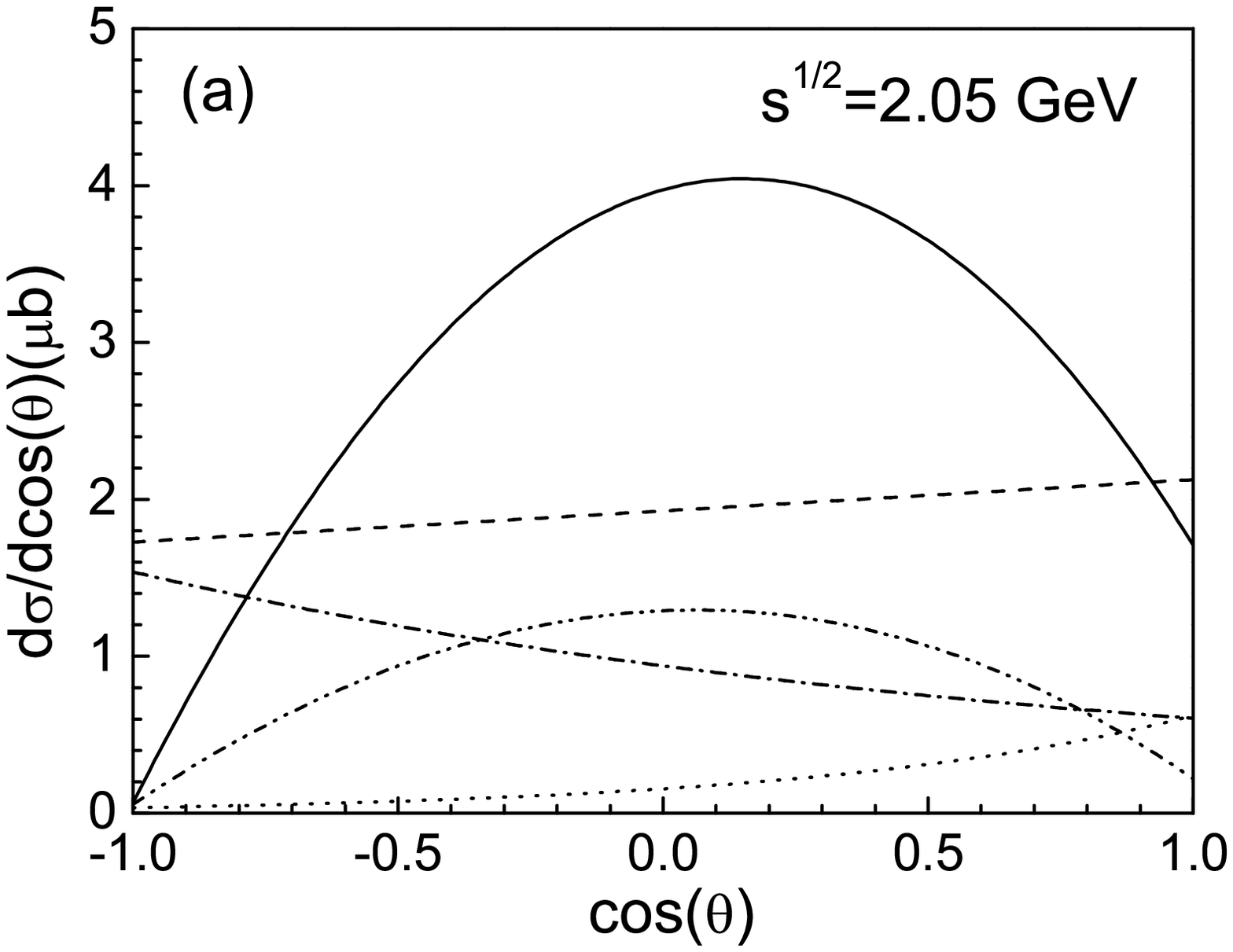}
\includegraphics[scale=0.4]{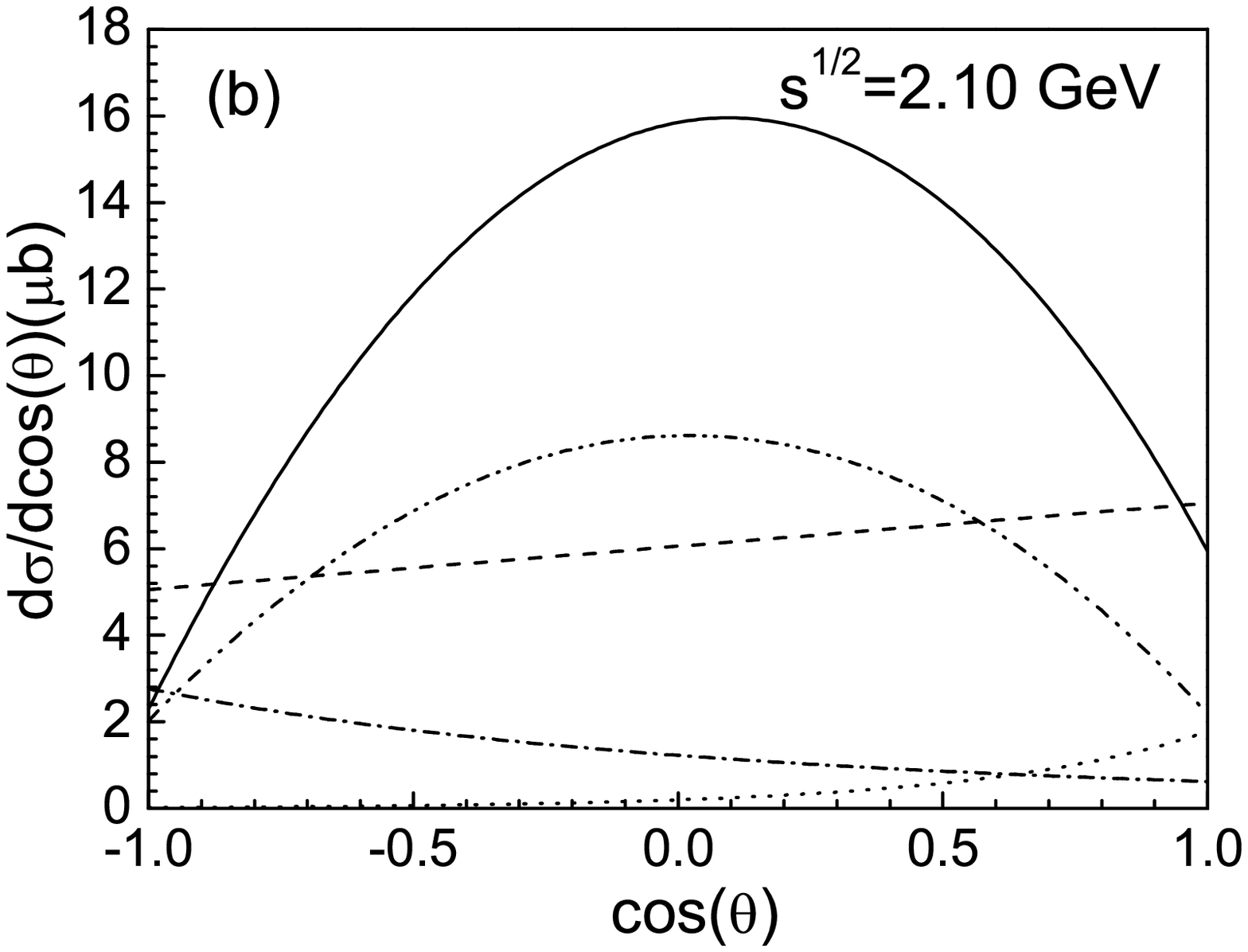}
\includegraphics[scale=0.4]{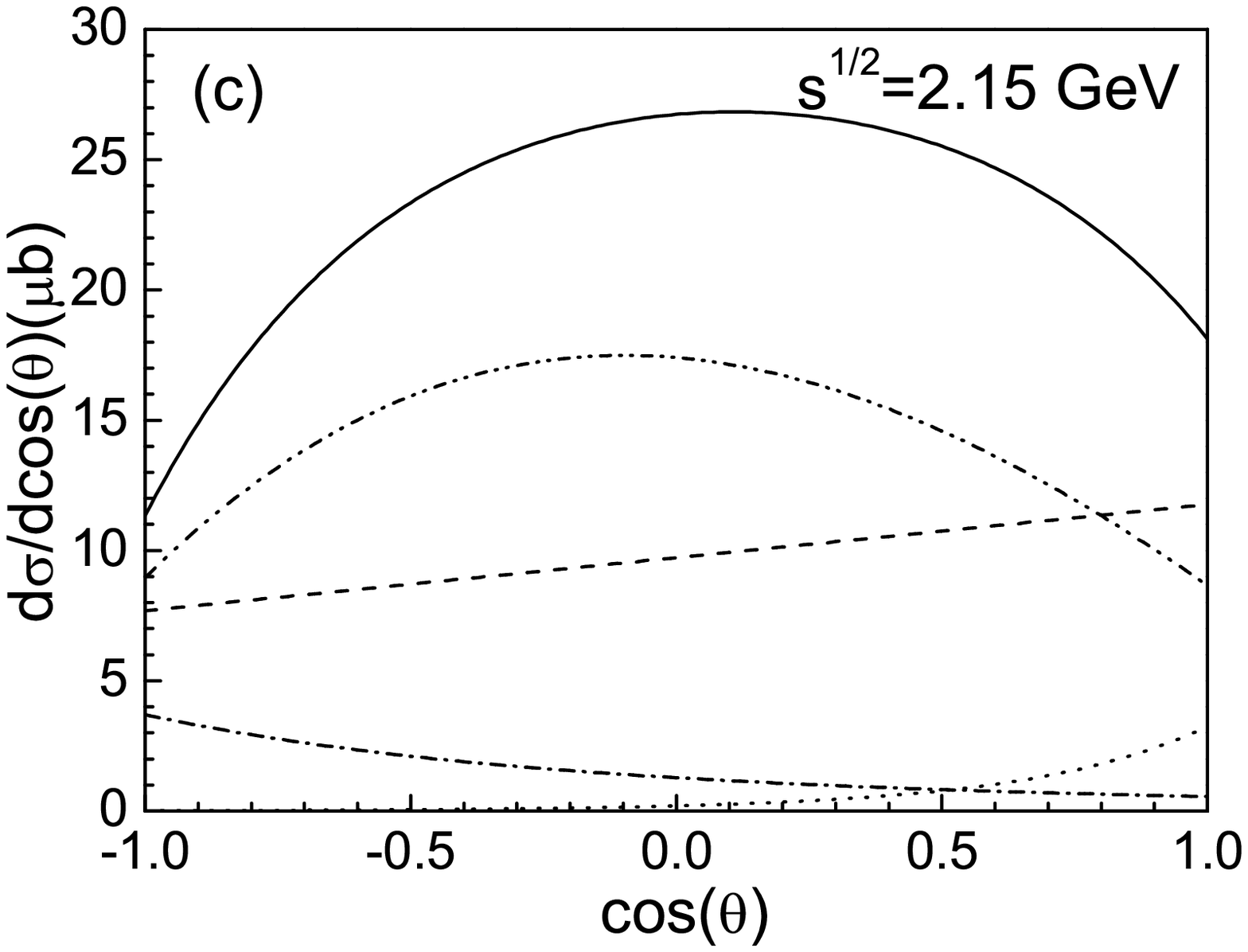}
\caption{Differential cross sections for $\pi^- p \to K^0
\Lambda(1520)$ reaction. The curves are the contributions from
$s-$channel nucleon pole term (dashed), $t-$channel $K^*$ term
(dotted), $u-$channel $\Sigma^+$ term (dash-dotted), $s-$channel
$N^*(2080)$ term (dash-dot-dotted) and the total contributions of
them (solid).} \label{pipdcs}
\end{center}
\end{figure}

With the above fitted parameters, the corresponding calculation
results for the differential cross sections for $\pi^- p \to K^0
\Lambda(1520)$ reaction at the energy around the central mass of
$N^*(2080)$ resonance, $\sqrt{s}=2.05$ GeV, $\sqrt{s}=2.10$ GeV, and
$\sqrt{s}=2.15$ GeV, are shown in Fig~\ref{pipdcs}(a),
Fig~\ref{pipdcs}(b), and Fig~\ref{pipdcs}(c), respectively. These
predictions can be checked by the future experiments.

Besides, with the strong coupling constant, $g_{N^*N\pi}$, which was
obtained from the $\chi^2-$fits, we have evaluated the $N^*(2080)$
resonnance to $N\pi$ partial decay width,
\begin{eqnarray}
\Gamma_{N^* \to N\pi} &=& \frac{g^2_{N^*N\pi}}{4\pi} \frac{|\vec{p}_N^{\,\,\rm c.m.}|^3}{m^2_{\pi}M_{N^*}}(E_N-m_N),
\end{eqnarray}
as deduced from the Lagrangian density of Eq.~(\ref{pinnstar}). In the above expression,
\begin{eqnarray}
E_N & =& \frac{M^2_{N^*}+m^2_N-m^2_{\pi}}{2M_{N^*}}, \\
|\vec{p}_N^{\,\,\rm c.m.}| &=& \sqrt{E^2_N-m^2_N}.
\end{eqnarray}

With the values of $M_{N^*} = 2115$ MeV, $\Gamma_{N^*} = 254$ MeV,
and also $g_{N^*N\pi} = 0.14 \pm 0.04$, we can get
\begin{eqnarray}
{\rm Br} (N^* \to N\pi) = \frac{\Gamma_{N^* \to N\pi}}{\Gamma_{N^*}} = (2.9 \pm 1.6) \%,
\end{eqnarray}
with the error from the uncertainty of the coupling constant
$g_{N^*N\pi}$. The above value is consistent with the result $(6 \pm
2) \%$ that was obtained from the multichannel
analysis~\cite{anisovich}.

\section{Numerical results for $pp \to pK^+ \Lambda(1520)$ reaction and discussions}

With the formalism and ingredients given above, the calculations of
the differential and total cross sections for $pp \to pK^+
\Lambda(1520)$ are straightforward,
\begin{eqnarray}
&& d\sigma (pp\to pK^+ \Lambda(1520)) = \frac{1}{4}\frac{m^2_p}{F} \sum_{s_1, s_2} \sum_{s_3, s_4} |{\cal M}|^2 \times \nonumber \\
&& \frac{m_p d^{3} p_{3}}{E_{3}} \frac{m_{\Lambda(1520)} d^{3} p_4}{E_4} \frac{d^{3} p_5}{2 E_5} \delta^4 (p_{1}+p_{2}-p_{3}-p_{4}-p_5), \nonumber \\
\label{ppdcs}
\end{eqnarray}
with the flux factor
\begin{eqnarray}
F=(2 \pi)^5\sqrt{(p_1\cdot p_2)^2-m^4_p}~. \label{eqff}
\end{eqnarray}

\begin{figure}[htbp]
\begin{center}
\includegraphics[scale=0.45]{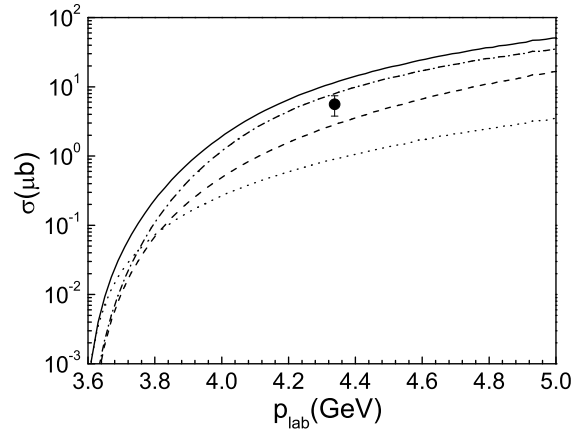}
\caption{Total cross sections vs beam energy ${\rm p_{lab}}$ of
proton for the $pp \to pK^+\Lambda(1520)$ reaction from present
calculation. The dashed, dotted, and dash-dotted lines stand for
contributions from nucleon pole, $\Sigma^+$ pole and $N^*(2080)$
resonance, respectively. Their total contribution are shown by the
solid line. The one experimental data point is taken from
Ref.~\cite{l1520data}.} \label{pptcs}
\end{center}
\end{figure}

The total cross section versus the beam energy (${\rm p_{lab}}$) of
the proton for the $pp \to p K^+ \Lambda(1520)$ reaction is
calculated by using a Monte Carlo multi-particle phase space
integration program. The results for beam energies ${\rm p_{lab}}$
from just above the production threshold $3.59$ GeV to $5.0$ GeV are
shown in Fig.~\ref{pptcs}. The dashed, dotted, and dash-dotted lines
stand for contributions from nucleon pole, $\Sigma^+$ pole and
$N^*(2080)$ resonance, respectively. Their total contributions are
shown by the solid line.~\footnote{Since the $t-$channel $K^*$ meson
exchange gives very small contribution to the $\pi^- p \to K^0
\Lambda(1520)$ reaction, especially for the invariant mass of
$K\Lambda(1520)$ below $2.4$ GeV, so in the calculation for $pp \to
pK^+\Lambda(1520)$ reaction, we ignore its contribution.} From
Fig.~\ref{pptcs}, we can see that the contribution from the
$u-$channel $\Sigma^+$ exchange is predominant at the very close to
threshold region, but, when the beam energy goes up, the
contributions from $s-$channel nucleon pole and $N^*(2080)$
resonance turn to be very important.

\begin{figure*}[htbp]
\begin{center}
\includegraphics[scale=0.8]{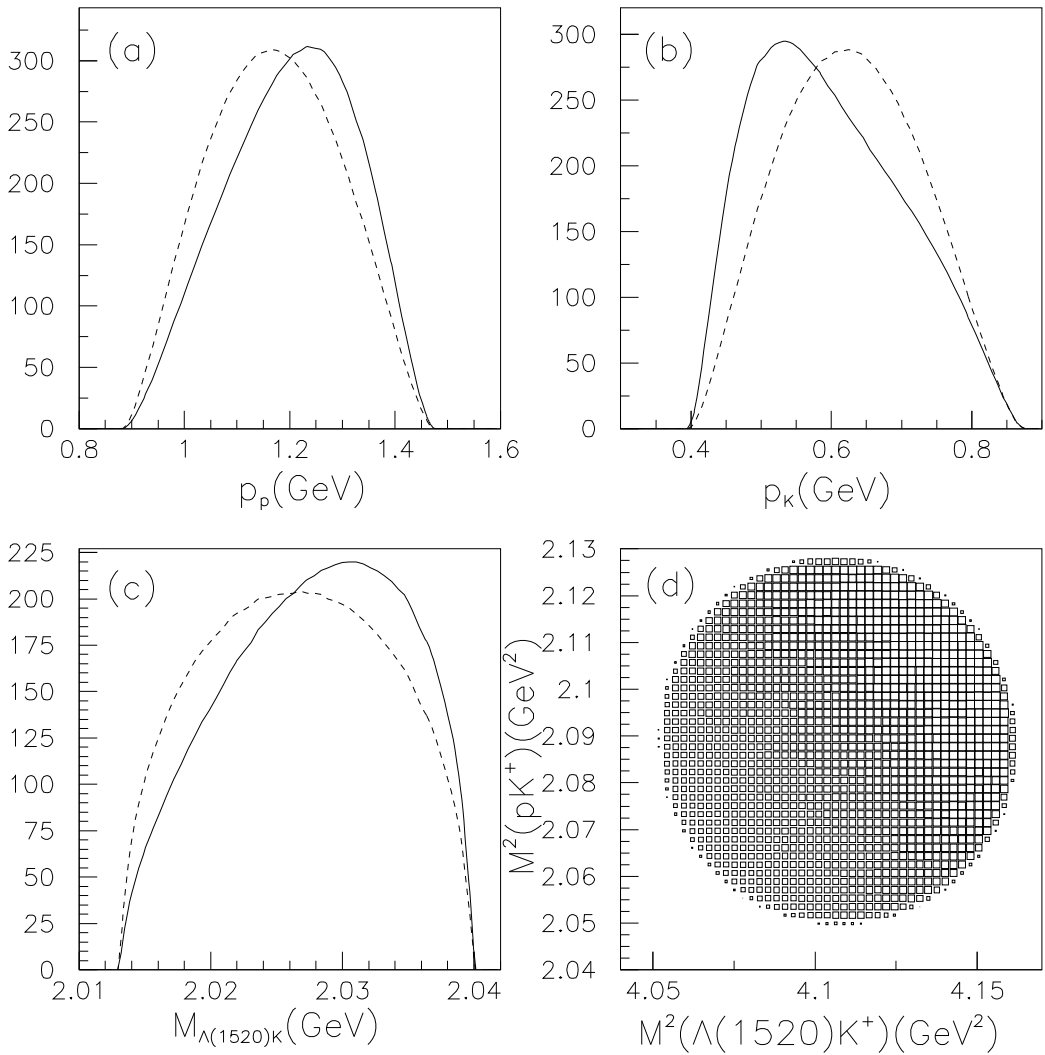}
\caption{Momentum distribution (arbitrary units), invariant mass
spectrum (arbitrary units), and Dalitz Plot for the $pp \to pK^+
\Lambda(1520)$ reaction at beam energy ${\rm p_{lab}} = 3.67$ GeV
comparing with the phase space distribution.The dashed lines are
pure phase space distributions, while the solid lines are full
results from our model.} \label{plab367}
\end{center}
\end{figure*}

It is important to note that our predictions for the total cross
section of $pp \to pK^+\Lambda(1520)$ reaction, at ${\rm p_{lab}} =
3.65$ GeV, is $0.01 \mu b$, which is $20$ times smaller than the
experimental upper limit $0.2 \mu b$ as measured by the COSY-ANKE
Collaboration~\cite{zychorcosy}. This shows that our model
predictions are consistent with the experimental results. Moreover,
the total cross section of $pp \to pK^+\Lambda(1520)$ reaction is
measured with HADES~\cite{l1520data} at GSI at kinetic beam energy
${\rm T_p} = 3.5$ GeV (corresponding to ${\rm p_{lab}} = 4.34$
GeV)\footnote{${\rm p_{lab}}$ = $\sqrt{{\rm E_{lab}}^2 - m^2_p}$ =
$\sqrt{({\rm T_p}+m_p)^2-m^2_p}$.}. The result is
$5.6\pm1.1\pm0.4^{+1.1}_{-1.6} ~\mu$b, as shown in Fig.~\ref{pptcs},
by comparing with our theoretical result, $11.5 ~\mu$b. If we modify
the cut off parameters $\Lambda_{\pi}$ and $\Lambda^*_{\pi}$ from
1.3 GeV to 1.0 GeV, we get $\sigma = 5.45 ~\mu$b, which is in
agreement with the experimental data well. However, it does not make
sense to fit the only one data point. So we still keep
$\Lambda_{\pi} = \Lambda^*_{\pi}$ = 1.3 GeV as used in many previous
works~\cite{xiezou}. We should also mention that, in the present
calculation, we did not include the $\Lambda(1520) p$
final-state-interaction (FSI), which can increase the results even
by a factor of 10 at the very near threshold region, such as the
important role played by $\Lambda p$ FSI in the $pp \to pK^+\Lambda$
reaction~\cite{Xie:2011me}. This is because there are no
experimental data on this reaction and also very scarce information
about the $\Lambda(1520) p$ FSI.

Furthermore, the corresponding momentum distribution~\footnote{It is
worth to note that our results are calculated in the reaction
laboratory frame, in which the target proton is at rest.} of the
final proton and $K^+$ meson, the $K\Lambda(1520)$ invariant mass
spectrum, and also the Dalitz Plot for the $pp \to pK^+
\Lambda(1520)$ reaction at beam momentum ${\rm p_{lab}} = 3.67$ GeV,
which is accessible for DISTO Collaboration~\cite{baledisto}, are
calculated and shown in Fig.~\ref{plab367}(a),
Fig.~\ref{plab367}(b), Fig.~\ref{plab367}(c), and
Fig.~\ref{plab367}(d), respectively. The dashed lines are pure phase
space distributions, while, the solid lines are full results from
our model. From Fig.~\ref{plab367}, we can see that even at ${\rm
p_{lab}} = 3.67$ GeV, there is a clear bump in the $K\Lambda(1520)$
invariant mass distribution, which is produced by including the
contribution from $N^*(2080)$ resonance.

\begin{figure*}[htbp]
\includegraphics[scale=0.8]{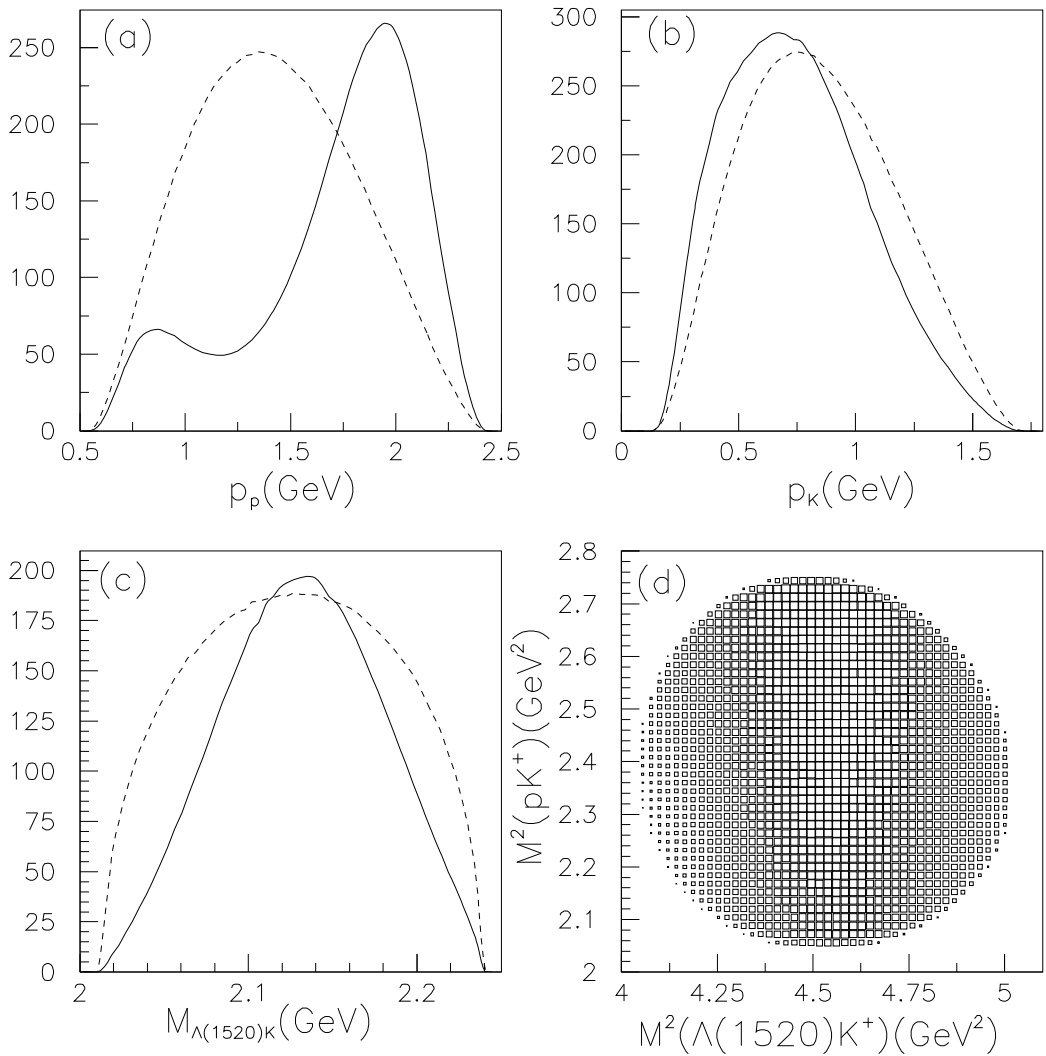}
\vspace{-0.2cm} \caption{As in Fig.~\ref{plab367}, but for ${\rm
p_{lab}} = 4.34$ GeV.} \label{plab42}
\end{figure*}

At the energy point of beam momentum ${\rm p_{lab}} = 3.67$ GeV, the
contribution from $u-$channel $\Sigma^+$ exchange is still dominant,
so, for comparing, we also present our calculated differential
distributions at ${\rm p_{lab}} = 4.34$ GeV where the contribution
from the $s-$channel nucleon pole and $N^*(2080)$ resonance is
dominant. Our results are shown in Fig.~\ref{plab42}. We can see
that our model results for the momentum distribution of final proton
are much different from the phase space distribution.

The momentum distribution, invariant mass spectra and the Dalitz
plots in Figs.~\ref{plab367} and~\ref{plab42} show direct
information about the $pp \to pK^+\Lambda(1520)$ reaction mechanism
and may be tested by the future experiments.

\section{Summary}

In this paper, the $\Lambda(1520)$ hadronic production in
proton-proton and $\pi^- p$ collisions are studied within the
combination of the effective Lagrangian approach and the isobar
model. For $\pi^- p \to K^0 \Lambda(1520)$ reaction, in addition to
the "background" contributions from $t-$channel $K^*$ exchange,
$u-$channel $\Sigma^+$ exchange, and $s-$channel nucleon pole terms,
we also considered the contribution from the nucleon resonance
$N^*(2080)$ (spin-parity $J^P = 3/2^-$), which has significant
coupling to $K\Lambda(1520)$ channel. We show that the inclusion of
the nucleon resonance $N^*(2080)$ leads to a fairly good description
of the low energy experimental total cross section data of $\pi^- p
\to K^0 \Lambda(1520)$ reaction. The $s-$channel nucleon pole and
$N^*(2080)$ resonance and also the $u-$channel $\Sigma^+$ exchange
give the dominant contributions below invariant mass $\sqrt{s}=2.4$
GeV, while the $t-$channel $K^*$ exchange diagram gives the minor
contribution.

From $\chi^2-$fits to the available experimental data for the $\pi^-
p \to K^0 \Lambda(1520)$ reaction, we get the $N^*(2080)N\pi$
coupling constant $g_{N^*(2080)N \pi} = 0.14 \pm 0.04$, which gives
the branching ration of $N^*(2080)$ resonance to $N\pi$ as $(2.9 \pm
1.6) \%$. Our result is consistent with the previous work. Besides,
the corresponding predictions for the differential cross sections of
$\pi^- p \to K^0\Lambda(1520)$ are also shown, by which the future
experiments can check our model.

Basing on the study of $\pi^-p \to pK^+\Lambda(1520)$ reaction, we
study the $pp \to pK^+\Lambda(1520)$ reaction with the assumption
that the production mechanism is due to the $\pi^0$-meson exchanges.
We give our predictions about total cross sections of this reaction.
Our results show that the contribution from the $u-$channel
$\Sigma^+$ exchange is predominant at the very near threshold
region, but, when the beam energy goes up, the contributions from
$s-$channel nucleon pole and $N^*(2080)$ resonance turn to be very
important. Furthermore, we also demonstrate that the invariant mass
distribution and the Dalitz Plot provide direct information of the
$pp \to pK^+\Lambda(1520)$ reaction mechanisms and may be tested by
the future experiments.

Finally, we would like to stress that due to the important role
played by the resonant contribution in the $\pi^- p \to K^0
\Lambda(1520)$ and $pp \to pK^+\Lambda(1520)$ reactions, accurate
data for these reactions can be used to improve our knowledge on the
$N^*(2080)$ properties, which are at present poorly known. This work
constitutes a first step in this direction.

\section*{Acknowledgments}

We would like to thank Prof. Bing-Song Zou for useful discussions
and the CAS Theoretical Physics Center for Science Facilities for
support and hospitality during the initiation of this work. This
work is partly supported by the National Natural Science Foundation
of China under grants  11105126 and 10905046.


\begin{thebibliography}{99}
%
\bibitem{klempt} E. Klempt and J. M. Richard, Rev. Mod. Phys. \textbf{82}, 1095 (2010).
%
\bibitem{pdg2012} J. Beringer et al., [Particel Data Group], Phys. Rev. D \textbf{86}, 010001 (2012).
%
\bibitem{capstick2000}S. Capstick and W. Robert, Prog. Part. Nucl. Phys. \textbf{45}, S241 (2000), and references therein.
\bibitem{Zou:2008be}
  B.~S.~Zou,
  eConf C {\bf 070910}, 112 (2007), and references therein.  
%
\bibitem{baledisto}F. Balestra et al., Phys. Rev. C \textbf{63}, 024004 (2001).
\bibitem{Quentmeier:2001ec}
  C.~Quentmeier, H.~H.~Adam, J.~T.~Balewski, A.~Budzanowski, D.~Grzonka, L.~Jarczyk, A.~Khoukaz and K.~Kilian {\it et al.},
  Phys.\ Lett.\ B {\bf 515}, 276 (2001). 
\bibitem{Winter:2006vd}
  P.~Winter, M.~Wolke, H.~-H.~Adam, A.~Budzanowski, R.~Czyzykiewicz, D.~Grzonka, M.~Janusz and L.~Jarczyk {\it et al.},
  Phys.\ Lett.\ B {\bf 635}, 23 (2006). 

%
\bibitem{hartcosy}M. Hartmann et al., Phys. Rev. Lett. \textbf{96}, 242301 (2006).
%
\bibitem{zychorcosy}I. Zychor et al., Phys. Lett. B \textbf{660}, 167 (2008).
%
\bibitem{maedacosy}Y. Maeda et al., Phys. Rev. C \textbf{77}, 015204 (2008).
%
\bibitem{geng1405}L. S. Geng and E. Oset, Eur. Phys. J. A \textbf{34}, 405 (2007).
%
\bibitem{xiecolin1405}J. J. Xie and C. Wilkin, Phys. Rev. C \textbf{82}, 025210 (2010).
%

\bibitem{Roca:2006pu}
  L.~Roca, C.~Hanhart, E.~Oset and U.~-G.~Meissner,
  Eur.\ Phys.\ J.\ A {\bf 27}, 373 (2006). 


\bibitem{Paryev:2010zza}
  E.~Y.~Paryev,
  J.\ Phys.\ G {\bf 37}, 105101 (2010).  

\bibitem{xiejuan}J. J. Xie and J. Nieves, Phys. Rev. C \textbf{82}, 045205 (2010).
%
\bibitem{hejun}J. He and X. R. Chen, Phys. Rev. C \textbf{86}, 035204 (2012).
%
\bibitem{nam2013} Seung-il Nam, arXiv: 1212.6114.


\bibitem{Kamano:2009im}
  H.~Kamano, B.~Julia-Diaz, T.~-S.~H.~Lee, A.~Matsuyama and T.~Sato,
  Phys.\ Rev.\ C {\bf 80}, 065203 (2009). 

\bibitem{Suzuki:2009nj}
  N.~Suzuki, B.~Julia-Diaz, H.~Kamano, T.~-S.~H.~Lee, A.~Matsuyama and T.~Sato,
  Phys.\ Rev.\ Lett.\  {\bf 104}, 042302 (2010). 

\bibitem{Kamano:2008gr}
  H.~Kamano, B.~Julia-Diaz, T.~-S.~H.~Lee, A.~Matsuyama and T.~Sato,
  Phys.\ Rev.\ C {\bf 79}, 025206 (2009). 


\bibitem{Kamano:2011ih}
  H.~Kamano, S.~X.~Nakamura, T.~S.~H.~Lee and T.~Sato,
  Phys.\ Rev.\ D {\bf 84}, 114019 (2011). 


%
\bibitem{toki}H.~Toki, C.~Garc\'\i a-Recio, and J.~Nieves, Phys.\
Rev.\ D {\bf 77}, 034001 (2008).
%
\bibitem{wufq} F. Q. Wu, B. S. Zou, L. Li and D. V. Bugg, Nucl. Phys. A \textbf{735}, 111 (2004); F. Q. Wu, and B. S. Zou, Phys. Rev. D \textbf{73}, 114008 (2006).
%
\bibitem{Mosel} T. Feuster and U. Mosel, Phys. Rev. C \textbf{58}, 457 (1998); Phys. Rev. C \textbf{59}, 460 (1999);
G. Penner and U. Mosel, Phys. Rev. C \textbf{66}, 055211 (2002); ibid. C \textbf{66}, 055212 (2002);
V. Shklyar, H. Lenske and U. Mosel, Phys. Rev. C \textbf{72}, 015210 (2005).
%
%
\bibitem{nam2} S.~I.~Nam, Phys.\ Rev.\ C {\bf 81}, 015201 (2010).
%
\bibitem{zouprc03} B.S. Zou and F. Hussain, Phys. Rev. C \textbf{67}, 015204 (2003).
%
\bibitem{tsushima} K. Tsushima, S.W. Huang and A. Faessler, Phys. Lett. B \textbf{337}, 245 (1994); K. Tsushima, A. Sibirtsev and A.W. Thomas, Phys. Lett. B \textbf{39}, 29 (1997); K. Tsushima, A. Sibirtsev, A.W. Thomas and G.Q. Li,  Phys. Rev. C \textbf{59}, 369 (1999), Erratum-ibid. C \textbf{61}, 029903 (2000).

\bibitem {sibi} A. Sibirtsev and W. Cassing, nucl-th/9802019; A. Sibirtsev, K. Tsushima, W. Cassing and A. W. Thomas, Nucl. Phys. A \textbf{646}, 427 (1999).
%
\bibitem{Saghai} B. Julia-Diaz, B. Saghai, T.~S.~H. Lee and F. Tabakin, Phys.
Rev. C {\bf 73}, 055204 (2006).
%
\bibitem{xiezou}J.~-J.~Xie, B.~-S.~Zou and B.~-C.~Liu,
Chin.\ Phys.\ Lett.\  {\bf 22}, 2215 (2005);  
J.~-J.~Xie and B.~-S.~Zou,
  Phys.\ Lett.\ B {\bf 649}, 405 (2007); 
J.~-J.~Xie, B.~-S.~Zou and H.~-C.~Chiang,
  Phys.\ Rev.\ C {\bf 77}, 015206 (2008); 

B.~-S.~Zou and J.~-J.~Xie,
  Int.\ J.\ Mod.\ Phys.\ E {\bf 17}, 1753 (2008). 
%
\bibitem{rarita}W.~Rarita and J.~Schwinger, Phys.\ Rev.\ {\bf 60}, 61
(1941).
%
\bibitem{nath}L.~M.~Nath, B.~Etemadi, and J.~D.~Kimel, Phys.\ Rev.\ D {\bf 3}, 2153 (1971).
%

\bibitem{ohta89} K.~Ohta, Phys.\ Rev.\ C {\bf 40} (1989) 1335.
%
\bibitem{haberzettl98} H.~Haberzettl, C.~Bennhold, T.~Mart, T.~Feuster, Phys.\ Rev.\ C {\bf 58}, R40 (1998).
%
\bibitem{Davidson:2001rk}
  R.~M.~Davidson and R.~Workman, Phys.\ Rev.\  C {\bf 63}, 025210 (2001).

\bibitem{Janssen:2001wk}
  S.~Janssen, J.~Ryckebusch, D.~Debruyne and T.~Van Cauteren,
  Phys.\ Rev.\  C {\bf 65}, 015201 (2002).
%
\bibitem{pipdata}A. Baldini, V. Flamino, W.G. Moorhead and D.R.O. Morrison, $Landolt$-$B${\"{o}}$rnstein$, $Numerical$ $Data$ $and$ $Functional$ $Relationships$
        $in$ $Science$ $an$ $Technology$, vol.\textbf{12}, ed. by H. Schopper, Springer-Verlag(1988), $Total$ $Cross$ $Sections$ $of$ $High$ $Energy$ $Particles$.
%
\bibitem{anisovich}A. V. Anisovich et al., Eur. Phys. J. A \textbf{8}, 15 (2012).
\bibitem{Xie:2011me}
  J.~-J.~Xie, H.~-X.~Chen and E.~Oset,
  Phys.\ Rev.\ C {\bf 84}, 034004 (2011). 

%
\bibitem{l1520data}G. Agakishiev {\it et al.}, [HADES Collaboration], arXiv: 1208.0205.

\end{thebibliography}
\end{document}